\documentclass[%
 aip,
 amsmath,amssymb,
 reprint,%
]{revtex4-1}

\usepackage{graphicx}
\usepackage{dcolumn}
\usepackage{bm}
\usepackage[columnwise]{lineno}
\usepackage{siunitx}
\usepackage{xcolor}
\usepackage[utf8]{inputenc}
\usepackage[T1]{fontenc}
\usepackage{mathptmx}
\usepackage[normalem]{ulem}

\makeatletter
\let\@fnsymbol\@fnsymbol@latex
\@booleanfalse\altaffilletter@sw
\makeatother

\begin{document}

\preprint{AIP/123-QED}

\title{Thermal boundary resistance predictions with non-equilibrium Green’s function and molecular dynamics simulations}

\author{Yuanchen Chu}
\altaffiliation{These authors contributed equally to this work, email: chu72@purdue.edu}
\affiliation{%
 School of Electrical and Computer Engineering, Purdue University, West Lafayette, IN 47907, USA\looseness=-1}%
 
\author{Jingjing Shi}
\altaffiliation{These authors contributed equally to this work, email: chu72@purdue.edu}
\affiliation{ 
School of Mechanical Engineering, Purdue University, West Lafayette, IN 47907, USA
}

\author{Kai Miao}
 \affiliation{%
 School of Electrical and Computer Engineering, Purdue University, West Lafayette, IN 47907, USA\looseness=-1}%

\author{Yang Zhong}
\affiliation{ 
School of Mechanical Engineering, Purdue University, West Lafayette, IN 47907, USA
}

\author{Prasad Sarangapani}
\affiliation{%
 School of Electrical and Computer Engineering, Purdue University, West Lafayette, IN 47907, USA\looseness=-1}%

\author{Timothy S. Fisher}
\affiliation{%
Department of Mechanical and Aerospace Engineering, University of California, Los Angeles, CA 90095, USA\looseness=-1
}%

\author{\\Gerhard Klimeck}
\affiliation{%
 School of Electrical and Computer Engineering, Purdue University, West Lafayette, IN 47907, USA\looseness=-1}%
\affiliation{%
 Network for Computational Nanotechnology, Purdue University, West Lafayette, Indiana 47907, USA\looseness=-1}%
\affiliation{
 Purdue Center for Predictive Materials and Devices, West Lafayette, Indiana 47907, USA\looseness=-1}

\author{Xiulin Ruan}
\affiliation{ 
School of Mechanical Engineering, Purdue University, West Lafayette, IN 47907, USA
}

\author{Tillmann Kubis}
\affiliation{%
 School of Electrical and Computer Engineering, Purdue University, West Lafayette, IN 47907, USA\looseness=-1}%
\affiliation{%
 Network for Computational Nanotechnology, Purdue University, West Lafayette, Indiana 47907, USA\looseness=-1}%
\affiliation{
 Purdue Center for Predictive Materials and Devices, West Lafayette, Indiana 47907, USA\looseness=-1}
\affiliation{
 Purdue Institute of Inflammation, Immunology and Infectious Disease, West Lafayette, Indiana 47907, USA\looseness=-1}

\date{\today}

\begin{abstract}
The non-equilibrium Green’s function (NEGF) method with B{\"u}ttiker probe scattering self-energies is assessed by comparing its predictions for the thermal boundary resistance with molecular dynamics (MD) simulations. 
For simplicity, the interface of Si/heavy-Si is considered, where heavy-Si differs from Si only in the mass value. 
With B{\"u}ttiker probe scattering parameters tuned against MD in homogeneous Si, the NEGF-predicted thermal boundary resistance quantitatively agrees with MD for wide mass ratios.
Artificial resistances that the unaltered Landauer approach yield at virtual interfaces in homogeneous systems are absent in the present NEGF approach. 
Spectral information result from NEGF in its natural representation without further transformations. 
The spectral results show that the scattering between different phonon modes plays a crucial role in thermal transport across interfaces.
B{\"u}ttiker probes provide an efficient and reliable way to include anharmonicity in phonon related NEGF.
NEGF including the B{\"u}ttiker probes can reliably predict phonon transport across interfaces and at finite temperatures.

\end{abstract}

\maketitle

Semiconductor nanodevices such as quantum cascade lasers, LEDs and thermoelectric devices are typically composed of several semiconductor materials~\cite{Alferov1998,Faist1994,Nakamura1994,Tsai2014}.
Scattering of thermal energy carriers at the interface between two materials results in thermal boundary resistance~\cite{Swartz1989}. 
The size of the thermal boundary resistance was previously reported~\cite{Landry2009} to be comparable to that of pure materials with lengths of a few to tens of nanometers.
Predicting the thermal boundary resistance gives important insight into the device physics and enables design improvements.
Often, molecular dynamics (MD) is used to model the thermal boundary resistance and reproduce experimental data~\cite{Schelling2002}. 
Inelastic phonon scattering is included in MD simulations through the anharmonicity of the interatomic potential~\cite{Hopkins2009}. 
The non-equilibrium Green’s function (NEGF) method~\cite{Datta2000} is widely accepted as one of the most consistent methods for electronic quantum transport in nanodevices~\cite{KNOCH2007572,Matyas2010}.
In particular for predicting stationary device physics, NEGF is potentially more attractive than MD given that it is a spectral approach when setup in energy space, though modal methods in non-equilibrium MD have just begun to be developed~\cite{PhysRevB.90.134312,PhysRevB.95.195202}.
When electrons and phonons are both solved in the NEGF framework, interparticle interactions and energy and momentum transfer in e.g. self-heating or thermoelectric situations can be described on equal footing with the predictions of the respective particles' propagation~\cite{Stieger2017}.
For phonon transport, however, the NEGF method has been used predominantly in the coherent (harmonic) regime due to the fact that the inclusion of incoherent scattering such as phonon-phonon decay usually requires solving polarization graphs in the self-consistent Born approximation which entails a large numerical load~\cite{PhysRevB.86.245407}.
It has been shown that the lack of anharmonicity in NEGF simulation gives incorrect thermal boundary resistance predictions at high temperatures.~\cite{Gaskins2018, Polanco2017, Le2017}.

In this work, a numerically efficient method to solve phonon transport in the NEGF framework including phenomenological phonon scattering with B{\"u}ttiker probes is presented and benchmarked against MD. 
The artificial resistance at the virtual interface in a homogeneous structure that plagues the equilibrium Landauer approach~\cite{Landry2009,shi2018dramatic} is absent in the presented NEGF approach. 
When solved for homogeneous systems, this NEGF method yields vanishing interface resistance. 
The thermal boundary resistance calculated with this NEGF method shows quantitative agreement with MD simulations.
The extracted spectral transport information from NEGF shows that the different phonon modal contributions play an important role in thermal transport across the interface.
\begin{figure}[h]
\centering
\includegraphics[width=0.46\textwidth,keepaspectratio]{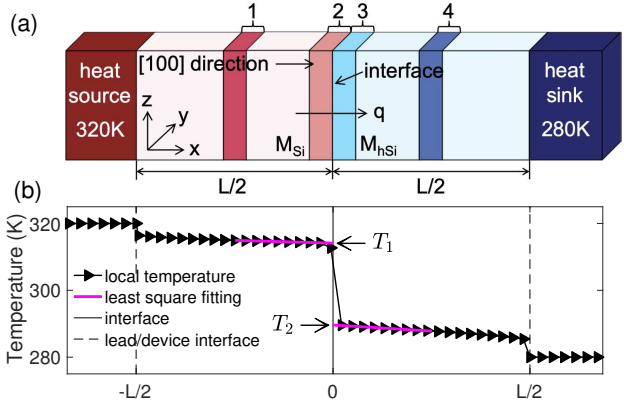}
\caption{\label{fig1}(a) Simulation domain considered in this work. Regions marked by $1$ through $4$ are three atomic layers in the middle of Si, left to the interface, right to the interface and in the middle of heavy Si, respectively. (b) Schematic of the thermal boundary resistance extraction.}
\end{figure}

Fig.~\ref{fig1}(a) shows the simulation domain considered in both MD and NEGF.
The system consists of Si to the left and heavy-Si to the right of an interface at position 0.
The heavy-Si differs from Si only in its atomic mass ratio vs. Si $MR=M_{hSi}/M_{Si}$.
For all simulations in this work, $M_{Si}$ is fixed at $\SI{28.085}{u}$ and a range of $1$ to $10$ is considered for $MR$.
Transport is solved within a range $L/2$ to the left and right of the interface.
The lattice temperature for regions further to the left and right of the interface is assumed constant and to equal $\SI{320}{\kelvin}$ and $\SI{280}{\kelvin}$, respectively.
Phonon transport occurs along the $x$ direction and the system is considered to be periodic along $y$ and $z$ directions.
The harmonic phonon bandstructure is described with a Tersoff potential~\cite{PhysRevB.39.5566} and the lattice constant is set to $\SI{5.431}{\angstrom}$.
Details of the thermal boundary resistance extraction are illustrated in Fig.~\ref{fig1}(b).
Firstly, linear fits are performed on the local temperature profiles to the left and to the right of the interface, respectively.
The local temperature $T$ is obtained by minimizing the difference between the local phonon energy density and the product of a local Bose-Einstein distribution and the local phonon density of states~\cite{Sadasivam2017}.
For a given atom, the local temperature $T$ solves the equation
\begin{equation}
\displaystyle\sum_{q_\parallel} \displaystyle \int_0^{\infty}\omega \rho(\omega)d\omega=\displaystyle\sum_{q_\parallel} \displaystyle \int_0^{\infty}\omega D(\omega) f_{BE}(\omega,T)d\omega,
\label{eq1}
\end{equation}
where $\omega$ is the phonon frequency and $q_\parallel$ is the transverse phonon wave vector. $\rho$ is the local phonon number density and $D$ is the local phonon density of states.
The temperatures $T_{1}$ and $T_{2}$ of the Si and heavy Si in the vicinity of the interface (see Fig.\ref{fig1}(b)) are determined with the fitted temperature profiles.
Finally, the thermal boundary resistance is calculated as $R=(T_{1}-T_{2})/q$, where $q$ is the simulated heat flux.
Following the discussion in Ref.~[\onlinecite{Sellan2010}], three different device lengths $L$ are simulated for each value of $MR$, and $R$ is extracted for the limit of $1/L=0$ by linear extrapolation.

The LAMMPS package~\cite{Plimpton1995} is used for all MD simulations in this work.
The lengths of both the heat source and the heat sink are $L/10$ and the simulation timestep is $\SI{0.4}{\femto\second}$.
Periodic boundary conditions are applied in $y$ and $z$ directions while the fixed boundary condition~\cite{doi:10.1063/1.1730376} is applied in the $x$ direction.
For MD calculations, the discretized device cross-section is of $8\times8$ conventional unit cells, with each conventional cell containing 8 atoms.
First, a canonical ensemble (NVT) is considered and run for $\SI{1.2}{\nano\second}$ to relax the structure, allowing the system to reach thermal equilibrium at $\SI{300}{\kelvin}$.
The system is then switched to a microcanonical ensemble (NVE) and a constant heat flux is added to the heat source and extracted from the heat sink for $\SI{12}{\nano\second}$.
After the system reaches steady state, the local temperature of each cell is obtained by averaging over ten million timesteps in the last $\SI{4}{\nano\second}$.
$L$ of $92$, $130$ and $184$ unit cells is used for MD based simulation of each $MR$ value.

\begin{figure}[h]
\centering
\includegraphics[width=0.483\textwidth,keepaspectratio]{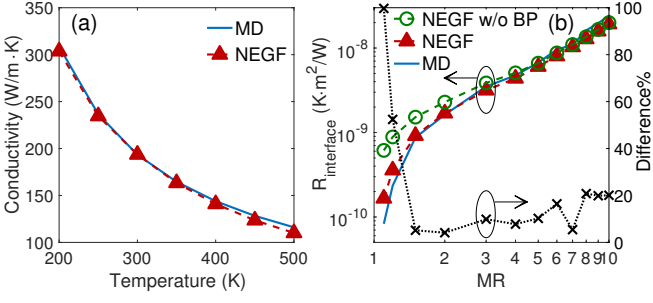}
\caption{\label{fig2}(a) Thermal conductivity of bulk Si calculated with MD and NEGF. The B{\"u}ttiker probe parameters $B$ and $C$ are fitted such that the transport results of NEGF agree with MD. (b) Thermal boundary resistance as a function of the mass ratio (MR) calculated by NEGF with the fitted parameters $B$ and $C$ of (a) and MD. 
The error bar of MD simulation is calculated but not plotted since the standard deviation for independent simulations is only \SI{1.6}{\percent} to \SI{4.6}{\percent} of the average resistance at the Si/heavy-Si interface for different mass ratios.
The dotted line is the relative difference of the two methods and is defined as  $(NEGF-MD)/MD\times100\%$.
The green curve shows the results calculated by NEGF without B{\"u}ttiker probe self-energies.
With increasing masses, the maximum phonon frequency and (with the frequency dependence of the Büttiker probes, Eq.~\ref{eq2}) the average scattering strength reduces. 
Consequently, the impact of scattering on $R_{interface}$ declines with increasing mass of heavy-Si.
}
\end{figure}

For all NEGF simulations, the nanodevice simulation tool NEMO5~\cite{6069914} is used.
Stationary Green's functions are solved in the energy domain, which gives spectral data without additional transformations.
To calculate the harmonic interatomic force constants (IFCs), a $3\times3\times3$ unit cell bulk Si structure is relaxed in LAMMPS using the Tersoff potential~\cite{PhysRevB.39.5566}.
The derivatives of the forces between atoms with respect to the atom position variations give the harmonic IFCs.
These values are then loaded into NEMO5 to construct the dynamical matrix~\cite{Paul2010}.
Only the transport direction ($x$) is discretized in real space.
The periodic directions ($y$ and $z$) are represented with a single conventional unit cell.
Longer-ranged periodicity is represented with the phonon momenta in reciprocal space.
Anharmonic phonon decay is included via B{\"u}ttiker probes~\cite{PhysRevLett.57.1761}.
The B{\"u}ttiker probe self-energies at atom $i$ with vibrational direction $m(x,y,z)$ are of the form 
\begin{equation}
\Sigma_{BP(i,m)}^{R}(\omega)=-i\frac{2\omega\hbar^{2}}{\tau_{(i,m)}(\omega)}.
\label{eq2}
\end{equation}
Each discretized atom in the system has a B{\"u}ttiker probe applied to it.
Following Ref.~[\onlinecite{Sadasivam2017}] we approximate the phonon frequency ($\omega$) dependent scattering lifetime $\tau$ as isotropic and assume it represents only the phonon-phonon Umklapp process~\cite{Miao2016}
\begin{equation}
\tau_{(i,m)}^{-1}(\omega)=\tau_{i}^{-1}(\omega)=BT_{i}\omega^{2}e^{-C/T_{i}}.
\label{eq3}
\end{equation}
$T_{i}$ is the phonon B{\"u}ttiker probe temperature of the atom $i$.
It represents the local temperature for the limit of complete phonon thermalization at each atom.
The actual local temperature differs from the B{\"u}ttiker probe one depending on the scattering strength~\cite{Miao2016}.
To ensure energy conservation, the B{\"u}ttiker probe temperature is solved iteratively by Newton's method until the integrated energy current vanishes for each B{\"u}ttiker probe~\cite{Sadasivam2017}.
Since the phonon Green's functions are solved with the recursive Green's function (RGF) method~\cite{Sadasivam2017}, the device is partitioned into slabs perpendicular to transport direction.
This limits the peak memory usage during computation, but requires the B{\"u}ttiker probe self-energies to be equal throughout each slab.
The parameters $B$ ($\SI{5e-20}{\second/\kelvin}$) and $C$ ($\SI{430}{\kelvin}$) are chosen such that the NEGF prediction of the bulk Si thermal conductivity agrees with the MD solution (see Fig.~\ref{fig2}(a)).
The same values of $B$ and $C$ are used in Si and heavy-Si.
In this way, the dependence of the scattering rate on the atomic mass and deviations of the anharmonicity near the interface from the one of the volume materials are ignored.
This is not a fundamental limitation of the B{\"u}ttiker probes.
Future work on other materials and interfaces might likely require a more detailed B{\"u}ttiker probe model.
$L$ of $146$, $184$ and $220$ unit cells is used for NEGF based simulation of each $MR$ value.

\begin{figure}[h]
\centering
\includegraphics[width=0.483\textwidth,keepaspectratio]{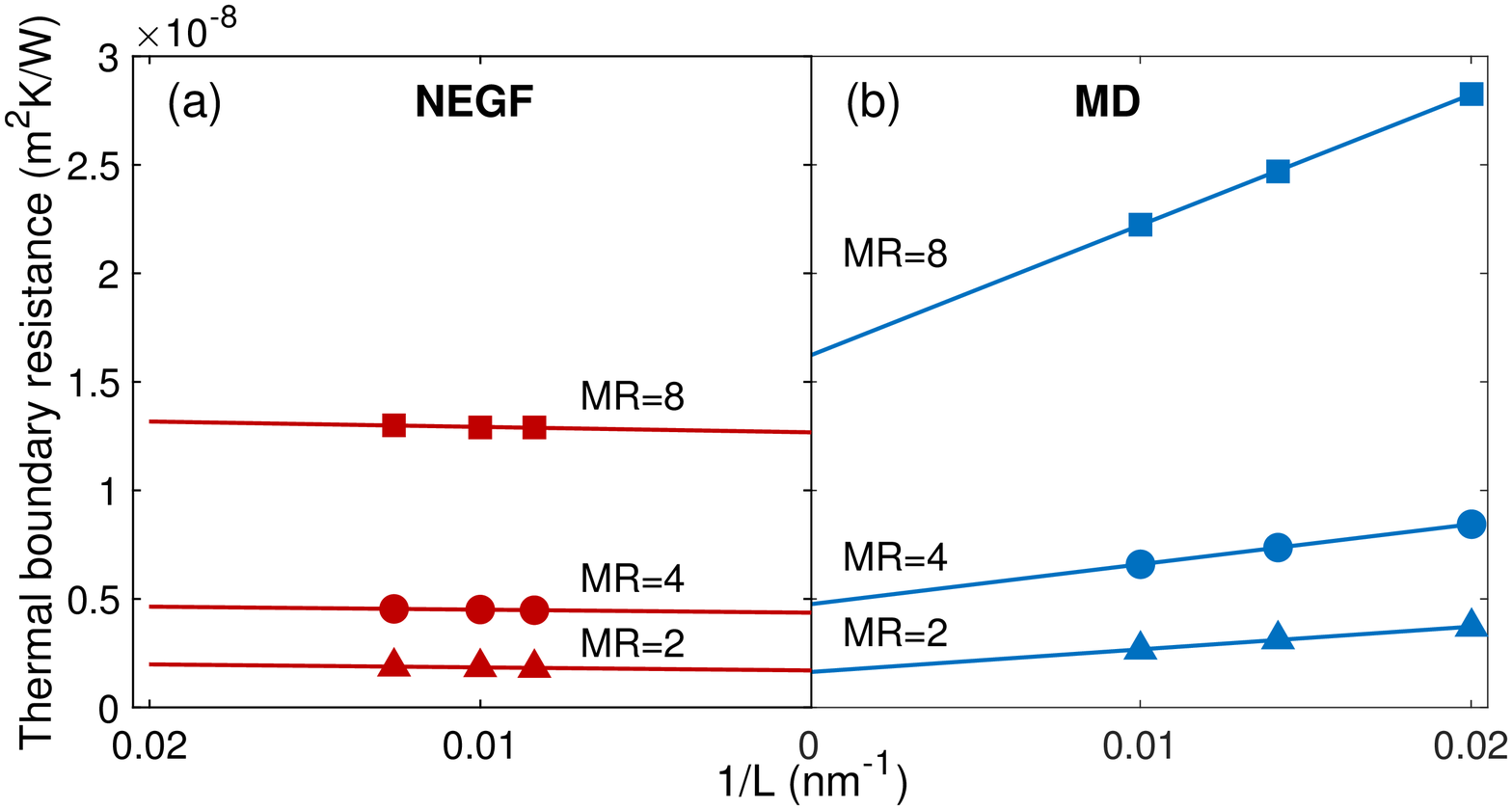}
\caption{\label{fig3}Linear extrapolation of (a) NEGF and (b) MD for three $MR$ values. MD results show a stronger dependence on the length of the device. Open boundary conditions included in NEGF with contact self-energies give almost device size independent results.}
\end{figure}

The spatial distribution of local phonon density of states (LDOS) can illustrate the contributions of different phonon modes to the heat flux in various regions.
For this purpose, four regions of interest, consisting of three atomic layers each are defined in Fig.~\ref{fig1}(a).
The LDOS $\phi_{i}(\omega)$ of any region $i$ of these regions is averaged over all of its atoms.
For a given phonon frequency, the L1-norm of the LDOS summed over all regions is defined as
\begin{equation}
\left\Vert\phi_{tot}(\omega)\right\Vert=\sum_{i=1}^{4}\phi_{i}(\omega).
\label{eq4}
\end{equation}
The relative contribution of LDOS of region $i$ is then defined as
\begin{equation}
R_{i}(\omega)=\frac{\phi_{i}(\omega)}{\left\Vert\phi_{tot}(\omega)\right\Vert}.
\label{eq5}
\end{equation}
Fig.~\ref{fig2}(b) shows the thermal boundary resistance as a function of $MR$ calculated with NEGF agrees quantitatively with the MD predictions.
For both methods, the thermal boundary resistance increases exponentially with $MR$ and vanishes when $MR$ tends to unity (i.e. for a homogeneous system).
Small remaining difference between the MD and NEGF results are systemic to the different treatment of phonon modes perpendicular to the transport direction:
In NEGF, the phonon momentum perpendicular to transport is explicitly resolved as a parameter in the Dyson and Keldysh equations~\cite{Sadasivam2017}. 
In contrast, MD calculations require as large as possible unit cells perpendicular to transport to cover as many phonon modes with long wave lengths in these directions as feasible.
Another source of differences can be the open boundary condition in transport direction:
Figs.~\ref{fig3} (a) and (b) benchmark the open boundary condition treatment in MD and NEGF, since they illustrate finite-size effects~\cite{Schelling2002} with the thermal boundary resistance $R$ as a function of the inverse system length $1/L$.  
For MD, the slope of $R$ vs $1/L$ increases with the mass ratio $MR$, i.e. with reducing average sound velocity in the device.
This agrees with similar findings discussed in detail in Ref.~[\onlinecite{Schelling2002}].
Note that the mean free path of Si at room temperature is about \SI{300}{\nano\meter}~\cite{Ju1999}. 
Therefore, it is practically impossible to completely eliminate the size effect along the transport direction in MD due to computational cost.
In contrast to the finite sized boundary reservoirs of MD, contact self-energies in NEGF incorporate semi-infinite leads as phonon reservoirs~\cite{datta_2005}.
Accordingly, we observe the slope of $R$ vs. $1/L$ in NEGF predictions is much smaller than that of MD.
Also its increase with $MR$ is comparably negligible.
This different boundary treatment is another source of some differences between MD and NEGF seen in Fig.~\ref{fig2} (b).

\begin{figure}[h]
\centering
\includegraphics[width=0.483\textwidth,keepaspectratio]{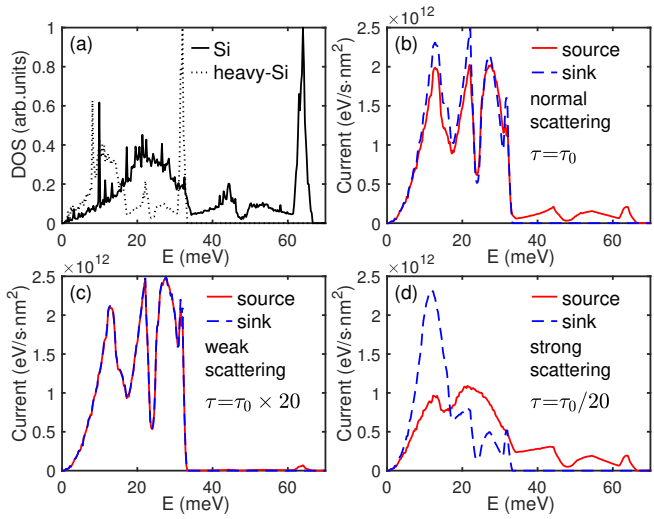}
\caption{\label{fig4}(a) Energy resolved (transverse momentum integrated) density of states in Si and heavy-Si leads. 
(b)$\sim$(d) Energy resolved (transverse momentum integrated) current of NEGF in the heat source and sink calculated with normal, artificially weak ($1/20\times$) and artificially strong ($20\times$) scattering strengths, respectively.
}
\end{figure}

Fig.~\ref{fig4}(a) shows the phonon density of states (DOS) in homogeneous Si and heavy-Si solved with NEGF.
The DOS of heavy-Si is limited to energies at or below $\SI{33.3}{\milli\electronvolt}$, two times less than in native Si in agreement with the applied $MR=4$.
Incoherent phonon scattering allows phonons with energies above this heavy-Si cutoff energy to propagate.
This is illustrated in Figs.~\ref{fig4}(b), (c) and (d) with the energy resolved current densities of the heat source and heat sink of the device in Fig.~\ref{fig1}(a) when solved in NEGF with different scattering strengths.
While the energy distribution changes with scattering, the total energy current (on the order of $\SI{e9}{\watt/\meter\squared}$) is conserved, and the difference of the total energy current between the source and the sink is on the order of $\SI{1}{\watt/\meter\squared}$.
In Fig.~\ref{fig4}(b), normal scattering strength (fitted to reproduce the MD-calculated bulk thermal conductivity) is applied.
The source current with energies above the energy cutoff is finite since its corresponding phonons can relax to lower energies at the heavy-Si side via inelastic scattering.
In Fig.~\ref{fig4}(c), an artificially weak scattering strength is used ($1/20\times$ normal scattering strength).
Accordingly, the profiles of current in both the heat source and sink follow the profile of the heavy-Si DOS. 
In contrast, Fig.~\ref{fig4}(d) shows the NEGF results when artificially strong scattering is used ($20\times$ normal scattering strength).
The current profiles in the heat source and heat sink follow the profiles of the DOS in the Si and the heavy-Si leads, respectively.
The results show that stronger inelastic scattering brings the system closer to local thermodynamic equilibrium.

\begin{figure}[h]
\centering
\includegraphics[width=0.38\textwidth,keepaspectratio]{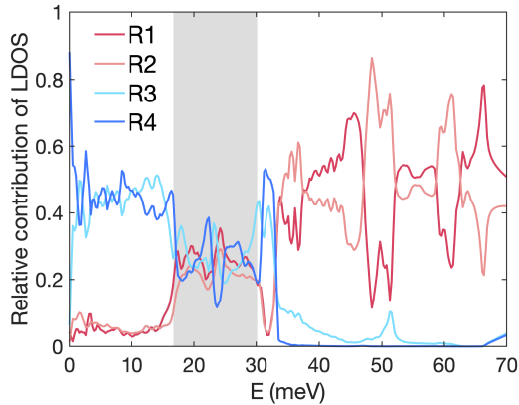}
\caption{\label{fig5}The ratio of the phonon mode in regions $1\sim4$ defined in Fig.~\ref{fig1} and by Eq.~\ref{eq4} ($\sum_{i=1}^{4}R_{i}=1$). The evenly distributed phonon modes around 22 meV (marked by the gray area) result in the highest current peak at $\SI{22}{\milli\electronvolt}$ (see Fig.~\ref{fig4}(b)) although having a lower DOS and lower group velocity compared to those at $\SI{12}{\milli\electronvolt}$.}
\end{figure}

In Fig.~\ref{fig4}(b), four current peaks in the heavy-Si heat sink are located at $\SI{12}{\milli\electronvolt}$, $\SI{22}{\milli\electronvolt}$, $\SI{27}{\milli\electronvolt}$ and $\SI{32}{\milli\electronvolt}$, respectively.
They correspond to the four peaks of the heavy-Si DOS shown in Fig.~\ref{fig4}(a).
The relative magnitudes of the four current peaks do not follow the relative magnitudes of the four DOS peaks (i.e., the list of current peaks arranged in decreasing order of magnitude is $\SI{22}{\milli\electronvolt}>\SI{12}{\milli\electronvolt}>\SI{27}{\milli\electronvolt}>\SI{32}{\milli\electronvolt}$, whereas the same list according to the DOS magnitude is $\SI{32}{\milli\electronvolt}>\SI{12}{\milli\electronvolt}>\SI{22}{\milli\electronvolt}>\SI{27}{\milli\electronvolt}$).
Without interfaces involved, the current is expected to be proportional to the product of the DOS and the group velocity~\cite{datta_1995}.
The results of Figs.~\ref{fig4} can be understood in Fig.~\ref{fig5}, since it illustrates the different relative contributions of phonon modes in different device areas.
Phonon modes around $\SI{12}{\milli\electronvolt}$ reside nearly exclusively in the heavy-Si.
Consequently, they contribute less to the overall heat current.
In contrast, phonon modes around $\SI{22}{\milli\electronvolt}$ are present in all 4 device regions considered in Fig.~\ref{fig5}.
Therefore, these modes can maintain a higher contribution to the total heat current. 

In conclusion, NEGF with B{\"u}ttiker probe scattering is applied to the thermal boundary resistance of the Si/heavy-Si interface.
The empirical NEGF scattering parameters are tuned to reproduce the thermal conductivity of homogeneous Si predicted by MD.
The scattering parameters proved to be transferable, since the NEGF results of the thermal boundary resistance quantitatively agrees with MD results for mass ratios ranging from $1$ to $10$.
The artificial resistance at the virtual interface in a homogeneous structure that plagues the equilibrium Landauer approach is absent in the presented NEGF approach. 
Besides, the present NEGF approach is found to be numerically more efficient than MD.
Thanks to the open boundary conditions, NEGF shows virtually no finite-size effects compared to MD.
The analysis of the NEGF spectral information shows that the scattering between different phonon modes determines the phonon energy current flow across interfaces.
Future improvements of our approach can include a more accurate description of inelastic scattering at the interface and enable computationally efficient calculation of spectral current adjacent to the interface.

Tillmann Kubis acknowledges support by Silvaco Inc. 
Jingjing Shi, Timothy Fisher and Xiulin Ruan would like to acknowledge the support by the Air Force Office of Scientific Research (AFOSR) MURI Grant (No. FA9550-12-1-0037).
This research is supported in part through computational resources provided by Information Technology at Purdue, West Lafayette, Indiana.
This research is part of the Blue Waters sustained-petascale computing project, which is supported by the National Science Foundation (Award No. ACI 1238993) and the state of Illinois.
Blue Waters is a joint effort of the University of Illinois at Urbana-Champaign and its National Center for Supercomputing Applications.


%

\end{document}